\documentclass[12pt]{article}

\usepackage{amsmath}
\usepackage{amsfonts}

\begin{document}

\title{Model of protein fragments and statistical potentials}

\author{S.V. Kozyrev\footnote{Steklov Mathematical Institute, the Russian Academy of Sciences}}

\maketitle

\begin{abstract}
We discuss a model of protein conformations where the conformations are combinations of short fragments from some small set. For these fragments we consider a distribution of frequencies of occurrence of pairs (sequence of amino acids, conformation), averaged over some balls in the spaces of sequences and conformations. These frequencies can be estimated due to smallness of the $\varepsilon$--entropy of the set of conformations of protein fragments.

We consider statistical potentials for protein fragments which describe the mentioned frequencies of occurrence and discuss model of free energy of a protein where the free energy is equal to a sum of statistical potentials of the fragments.

We also consider contribution of contacts of fragments to the energy of protein conformation, and contribution from statistical potentials of some hierarchical set of larger protein fragments. This set of fragments is constructed using the distribution of frequencies of occurrence of short fragments.

We discuss applications of this model to problem of prediction of the native conformation of a protein from its primary structure and to description of dynamics of a protein. Modification of structural alignment taking into account statistical potentials for protein fragments is considered and application to threading procedure for proteins is discussed.
\end{abstract}

\section{Introduction}

In the present paper we consider a model of protein free energy based on bionformatics. We construct a model of statistical (or empirical) potentials for short fragments of proteins. Application of this model to problem of prediction of the native conformation of a protein using primary structure of the protein, and to investigation of protein dynamics is discussed.

The main approach in physics is the construction of physical models starting from first principles (fundamental interactions). This approach is effective for systems with low complexity (the complexity can be understood as Kolmogorov complexity), but is less effective for complex systems, in particular in biology.

Application of computations from first principles to modeling of proteins results in a large amount of calculations, moreover, the computed dynamics will be unstable --- small perturbations of parameters of the model will result in large changes in behavior, not only the dynamics but also the lowest energy state may change.
Actually this property of proteins to have complex behavior which can be adjusted by changing their primary structures (sequences of amino acid residues in proteins) is crucially important for biology.

For complex systems we discuss the following approach --- instead of description from first principles one can consider a model which describes effective behavior of the system using a knowledge base about the system. The system will be described by some large database.  Our aim is to construct a big data physical model for conformations and dynamics of proteins.

Models of complex and disordered systems were widely investigated in theory of spin glasses \cite{MPV}. In this case the idea of genericity was used and the details of the disorder in the system (in particular the disorder matrix for the Sherrington--Kirkpatrick model) were not important. In this paper we will discuss how to take into account the particular properties of the disorder represented by some database.

For the investigation of proteins we will use a combination of models of fragments and statistical potentials. In models of statistical (empirical) potentials effective interactions in the system are reproduced using the empirical distribution functions, see  \cite{Finkelstein}, \cite{Rigden}. This approach is widely used in physics of proteins. For more discussion of physics of polymers see \cite{Finkelstein}, \cite{GrosbergKhokhlov}.

In the approach of fragments protein conformations are represented as combinations of short fragments (for instance  fragments may be of the length five amino acid residues) \cite{Unger,Micheletti,Brevern,Karplus,Kolodny,Jones,Rooman,Baker}, \cite{Nekrasov,Nekrasov1,Nekrasov2}. It was shown that with respect to some natural metrics the conformations of fragments can be clustered in small set of clusters of small radii. In the present paper we discuss this phenomenon as smallness of $\varepsilon$--entropy of the space of conformations of fragments. Lattice version of the model of fragments was discussed in \cite{quinary}, it was shown that in this model one obtains lattice models of protein secondary structures.

We construct a statistical potential for pairs (sequence of amino acids, conformation) for short fragments of proteins using averaging over some balls in the space of sequences. This statistical potential can be used for modeling of protein conformations. One can take into account the contacts of protein fragments in similar way. The obtained model of protein free energy will be a model with non--local cooperative interaction based on statistical potentials for protein fragments (an example of big data physical model).

The next component of the model is the hierarchical analysis of protein structure (a generalization of the approach of \cite{Nekrasov,Nekrasov1,Nekrasov2}). We consider distribution of frequencies of occurrence of short protein fragments as a function of a number of a fragment in the protein, and investigate the hierarchy of local maxima and minima for this function. We consider the contribution to the free energy of a protein from statistical potentials of the mentioned hierarchical set of structures. Hierarchical structure of polymer globules was also discussed for DNA packing \cite{GNS}.

The functional of free energy for this model will depend on approximately thousands of parameters. Most complex physical models have approximately dozens of parameters. This shows the difference of the degree of complexity between physical and minimally complex biological systems.

The exposition of the present paper is as follows. In section 2 we consider the model of fragments and in section 3 the construction of statistical potentials which describe joint distribution of pairs (sequence, conformation) for fragments of proteins. The construction of statistical potentials is based on the smallness of $\varepsilon$--entropy of the space of conformations of fragments. In section 4 we construct statistical potentials for contacts of fragments in analogous way. In section 5 we discuss hierarchical systems of larger fragments and statistical potentials for these systems.
In section 6 we discuss a model of combinatorial optimization for construction of native conformation of the protein from its primary structure, and consider a modification of the threading method using  statistical potentials of fragments. In section 7 we discuss the structural alignment construction related to the model of fragments and statistical potentials discussed in this paper.
In section 8 we discuss application to protein dynamics. Section 9 contains a conclusion of the paper.

\section{Space of fragments}

We consider the problem of correspondence between the primary structure and the native conformation of a protein. We use a variant of model of fragments of proteins (models of fragments were considered in many works, in particular in \cite{Unger,Micheletti,Brevern,Karplus,Kolodny,Jones,Rooman,Baker}). We consider a database $S$ --- a set of fragments of proteins. All fragments in $S$ have a fixed short length $N$ (for instance one can take $N$ equal to five amino acid residues). The database $S$ under consideration will contain pairs $(I,\Gamma)$ of the form (sequence of amino acids in the fragment, conformation of the fragment). We will describe the function $P(I,\Gamma)$ of joint distribution of sequences and conformations of fragments, using the database $S$. Then, using this function we will discuss native conformations of proteins.

Let us denote $X$ and $Y$ correspondingly the sets of all sequences of amino acid residues in fragments and conformations of fragments (not necessarily in $S$). Let us consider several metrics on these sets.

\medskip

\noindent{\bf Metric on the set of sequences of amino acids in fragments.}\quad
Let us consider the metric $d(\cdot,\cdot)$ on the space $X$ of sequences of amino acids in fragments of length $N$
\begin{equation}\label{metric_d}
d(I,J)=\sum_{l=1}^{N}\rho(A(i_l,j_l)).
\end{equation}
Here $I=i_1\dots i_N$, $J=j_1\dots j_N$ are sequences of amino acids ($i_l$, $j_l$ are the $l$-th amino acids in fragments $I$, $J$), $(A(i,j))$ is the matrix of some probabilistic model of evolution of proteins (PAM or BLOSUM matrices, see for example \cite{Pevzner}), i.e. the matrix element $A(i,j)$ is large when the probability of transition between amino acids $i$ and $j$ in the evolution model is large, $\rho$ is positive decreasing function.

This means that the distance between the fragments $I$, $J$ is equal to the sum of distances between amino acids in the corresponding positions, the distance between the amino acids is large when the probability of substitution of the amino acids in the model of evolution is small. Thus $d(\cdot,\cdot)$ measures the evolutionary distance between the fragments.

\medskip

\noindent{\bf Metrics on the set of conformations of fragments.}\quad
A metric on the space $Y$ of conformations of fragments can be introduced in different ways. One can parameterize the space $Y$ using sequences of pairs of dihedral angles $(\phi,\psi)$ (the Ramachandran diagrams) for the corresponding amino acids. In particular for fragments of length five the space $Y$ is the 8--dimensional torus (here we take into account the four inner vertices of the fragments and the corresponding dihedral angles).
For two conformations $\Gamma_1$, $\Gamma_2$ which correspond to sequences of dihedral angles $\{(\phi^{\alpha}_{1},\psi^{\alpha}_{1})\}$, $\{(\phi^{\alpha}_{2},\psi^{\alpha}_{2})\}$ we consider the metric of root mean square deviation
$$
s(\Gamma_1,\Gamma_2)=\sqrt{\sum_{\alpha=1}^{N-1} \left[ \left(\phi^{\alpha}_{1}-\phi^{\alpha}_{2}\right)^2 + \left(\psi^{\alpha}_{1}-\psi^{\alpha}_{2}\right)^2 \right] }.
$$

Another ways to introduce a metric are as follows
$$
\max_{\alpha=1}^{N-1}\sqrt{ \left(\phi^{\alpha}_{1}-\phi^{\alpha}_{2}\right)^2 + \left(\psi^{\alpha}_{1}-\psi^{\alpha}_{2}\right)^2  },
$$
$$
\sum_{\alpha=1}^{N-1}\sqrt{ \left(\phi^{\alpha}_{1}-\phi^{\alpha}_{2}\right)^2 + \left(\psi^{\alpha}_{1}-\psi^{\alpha}_{2}\right)^2  }.
$$

Metrics can be also defined in alternative way using representations of the conformations $\Gamma_1$, $\Gamma_2$ by sets of coordinates of $C_{\alpha}$--atoms of the fragments in the three dimensional space. The metric can be introduced by the expression
$$
s'(\Gamma_1,\Gamma_2)=\sqrt{\sum_{\alpha=0}^{N}\left((x^{\alpha}_1-x^{\alpha}_2)^2+(y^{\alpha}_1-y^{\alpha}_2)^2+(z^{\alpha}_1-z^{\alpha}_2)^2 \right)},
$$
where $(x^{\alpha}_1,y^{\alpha}_1,z^{\alpha}_1)$ are the coordinates of $C_{\alpha}$--atom in the conformation $\Gamma_1$ (analogously $(x^{\alpha}_2,y^{\alpha}_2,z^{\alpha}_2)$ for $\Gamma_2$). Here we choose the embeddings of conformations $\Gamma_1$, $\Gamma_2$ in $\mathbb{R}^3$ in such a way that the metric $s'$ is the infimum over possible embeddings of conformations. The summation runs over  $C_{\alpha}$--atoms (for a fragment of length $N$ one has $N+1$ such atoms).

Another possible metrics have the forms
$$
\max_{\alpha=0}^{N}\sqrt{(x^{\alpha}_1-x^{\alpha}_2)^2+(y^{\alpha}_1-y^{\alpha}_2)^2+(z^{\alpha}_1-z^{\alpha}_2)^2 },
$$
$$
\sum_{\alpha=0}^{N}\sqrt{(x^{\alpha}_1-x^{\alpha}_2)^2+(y^{\alpha}_1-y^{\alpha}_2)^2+(z^{\alpha}_1-z^{\alpha}_2)^2 }.
$$

\medskip

\noindent{\bf Coarse graining of the space of conformations of fragments.}\quad
It is known from the analysis of protein conformations in model of fragments \cite{Unger,Micheletti,Brevern,Karplus,Kolodny,Jones,Rooman,Baker} that the known conformations of fragments of proteins are concentrated in some sufficiently small subset of the space $Y$.
For the description of this phenomenon one can use a coarse graining procedure for observable conformations in $Y$ using a covering of the set $S\subset Y$ by some set of balls (with respect to some of the discussed above metrics in the space $Y$).

In particular \cite{Micheletti}, it was shown that all observable in experiments conformations in $S$ will belong to some covering of $S$ containing about one hundred of balls of small diameter $\varepsilon$ (approximately  1~\AA $~$ in metric $s'$ of root mean square deviation). Total number of balls of the same diameter in a covering of all space $Y$ is of orders of magnitude larger.

Therefore  {\it $\varepsilon$--entropy of observable in proteins subset of the space $Y$ of conformations of fragments is very small}.
Experimentally observable conformations of fragments are very specific if we ignore differences between conformations at small distances.

Taking into account this observation we denote $Y_{\varepsilon}$ the set of {\it coarse grained observed} conformations of fragments. The set $Y_{\varepsilon}$ can be understood as a covering of the set of conformations in $S$ by balls of the diameter $\varepsilon$ with respect to some of the described above metrics in $Y$, or equivalently this set can be considered as  $\varepsilon$--net for $S\subset Y$  (if we put in correspondence to a ball the center of this ball).

Usually when discussing $\varepsilon$--entropy one considers the dependence of the entropy on $\varepsilon$. In the case under consideration we are interested in the entropy for fixed  $\varepsilon$ (approximately 1~\AA ).

\medskip

\noindent{\bf $\varepsilon$--Entropy.}\quad Let us recall the definition of $\varepsilon$--entropy, see for instance \cite{Kolmogorov}.

Let $A$ be non--empty subset in a metric space $R$.

A system $\gamma$ of sets $U\subset R$ is called $\varepsilon$--covering of  $A$ if the diameter of any $U\in\gamma$ is less or equal $2\varepsilon$ and $A$ belongs to the union of $U\in\gamma$. We denote $N_{\varepsilon}(A)$ the minimal number of sets in a $\varepsilon$--covering of $A$.

Then $H_{\varepsilon}(A)=\log N_{\varepsilon}(A)$ is the $\varepsilon$--entropy of the set $A$.
\medskip

\noindent{\bf Protein conformations and sequences of fragments.}\quad For a protein with primary structure $I$ (the sequence of amino acid residues) and conformation $\Gamma$ we consider the corresponding set $(I_{i},\Gamma_{i})$ of fragments of length $N=5$ in the protein $(I,\Gamma)$, where $I_{i}$ is the sequence and $\Gamma_{i}\in Y_{\varepsilon}$ is the conformation of the $i$-th fragment in the protein (i.e the conformation is coarse grained in the discussed above sense).

Therefore a conformation $\Gamma$ of a protein can be represented by a sequence $(\Gamma_i)$ of symbols from $Y_{\varepsilon}$ which describe coarse grained conformations of fragments.

The described above procedure can not generate an arbitrary sequence $(\Gamma_i)$ of conformations of fragments starting from some protein conformation.
Conformations of neighbor fragments must be compatible, i.e. should have well defined intersection. For conformations $\Gamma_{i}$, $\Gamma_{i+1}$ from $Y_{\varepsilon}$ of the neighbor fragments the distance between the shorter fragments corresponding to the intersection of $\Gamma_{i}$, $\Gamma_{i+1}$ should be less or equal $\varepsilon$.

Let us consider the matrix $(C_{\Gamma\Delta})$, $\Gamma,\Delta\in Y_{\varepsilon}$ of intersections of fragments. The matrix element $C_{\Gamma\Delta}=1$ if the pair $(\Gamma,\Delta)$ of conformations of fragments is compatible, i.e. there exists a conformation of the fragment of the polymer of length six, where the first five residues have the conformation $\Gamma$ and the last five residues have the conformation $\Delta$. In the opposite case, if the pair $(\Gamma,\Delta)$ is incompatible, we put $C_{\Gamma\Delta}=0$.
This matrix nonsymmetric since a peptide chain is directed.

The matrix $(C_{\Gamma\Delta})$ is sparse, i.e. majority of matrix elements will be equal to zero. This matrix (for length of fragments $N=5$) has the dimension about $100\times 100$, and the number of non--zero matrix elements will be of order of thousand. Sparsity of this matrix puts considerable constraints on the size of the set of possible conformations of proteins. For the lattice version of model of fragments sparsity of this matrix implies lattice analogues of secondary structures \cite{quinary}.

\section{Statistical potentials for the model of fragments}

In order to construct joint distribution function of sequences of amino acids in fragments of proteins and conformations of the fragments starting from the database $S$ we have to perform averaging over sequences, because even after the coarse graining of conformations the database will be too small. We will use averaging over balls with respect to metric (\ref{metric_d}) in the space $X$ of sequences (i.e. use methods of nonparametric statistics).

We construct joint distribution function $P_{\delta,\varepsilon}(I,\Gamma)$, $I\in X$, $\Gamma\in Y_{\varepsilon}$
of pairs (sequence, conformation) from the database $S$ using averaging of the data $(I',\Gamma')\in S$ over the ball of diameter $\delta$ with center in $I$  with respect to metric $d(\cdot,\cdot)$ of the form (\ref{metric_d}) in the space $X$ of sequences. Here the conformation $\Gamma'$ should belong to the corresponding $\Gamma\in Y_{\varepsilon}$ element of the covering of the set of conformations of fragments and $d(I,I')\le\delta/2$. Therefore the function $P_{\delta,\varepsilon}(I,\Gamma)$ is equal to the number of fragments from $S$ which belong to the direct product of the ball in $X$ and the element $\Gamma\in Y_{\varepsilon}$ of the covering (i.e. $P_{\delta,\varepsilon}(I,\Gamma)$ is the coarse grained joint occurrence frequency for conformation and sequence of the fragment).

The diameter $\delta$ used in computation of $P_{\delta,\varepsilon}(I,\Gamma)$ is taken as follows: in average for fragments in a ball with center in $I$ for a fragment of length five three (or more) amino acids coincide with the corresponding amino acids in $I$ and two can differ.

A sample $S$ of the fragments is random, thus the function $P_{\delta,\varepsilon}(I,\Gamma)$ is also random. For sufficiently large samples $S$ and diameters $\delta$ this function should converge to a deterministic function.

\medskip

\noindent{\bf Statistical potentials.}\quad
For modeling of proteins one can use {\it statistical (or empirical) potentials}. The notion of a statistical potential is based on the idea that properties of proteins are distributed with respect to the Boltzmann distribution, see for example \cite{Finkelstein}, \cite{Rigden}. Let $p$ be some property of the protein (for instance some set of coordinates of relative positions of some residues). One can introduce the corresponding statistical potential by the formula
$$
E(p)=-\log n(p),
$$
where $n(p)$ is the observable value of occurrence in the database of the property $p$. Property $p$ may describe distances between $C_{\alpha}$--atoms of the backbone, values of dihedral angles etc.

In particular the Miyazawa--Jernigan matrix of energies of pairwise interactions of amino acids built using the statistics of contacts of amino acids in proteins is an example of statistical potential \cite{Miyazawa}.

In the above examples the statistical potentials were used for description of real interactions in proteins. We consider a different point of view --- we describe proteins by a set of convenient parameters (using the model of fragments), then define the energy of the model using statistical potentials for the parameters under consideration.

\medskip

\noindent{\bf Statistical potentials for the model of fragments.}\quad Let us consider the model of free energy of a protein, where the energy will be equal to a sum of contributions of fragments. For a fragment ($I$, $\Gamma$), $I\in X$, $\Gamma\in Y_{\varepsilon}$ we define energy of the pair $I$, $\Gamma$ by the coarse grained joint occurrence frequency as the statistical potential
\begin{equation}\label{Phi}
\Phi(I,\Gamma)=-\log P_{\delta,\varepsilon}(I,\Gamma).
\end{equation}

For a protein with the conformation $\Gamma$ and the primary structure $I$ we introduce the free energy functional
\begin{equation}\label{energy}
F_0(I,\Gamma)=\sum_{i}\Phi(I_{i},\Gamma_{i}),
\end{equation}
where the summation runs over fragments of length five in a protein, $I_{i}$ is the sequence and $\Gamma_{i}$ is the conformation of the $i$-th fragment of the protein ($I$, $\Gamma$). Physical meaning of this functional is described by the assumption that conformations of fragments with low energy are frequent in proteins (i.e. proteins are selected to make energies of native conformations to be low, equivalently, to make native conformations stable).

In particular,  $\Phi(I_{i},\Gamma_{i})$ should describe local properties (for example elasticity) of a protein at the $i$-th fragment. This idea can be used for comparison of different proteins and for the investigation of protein dynamics, see below.

\medskip

\noindent{\bf Remark.}\quad
Substitution of a small number of amino acid residues in a fragment $I$ by similar residues will not change dramatically the probabilities for this fragment to take conformations $\Gamma\in Y_{\varepsilon}$. Therefore the function $\Phi(I,\Gamma)$ should depend on $I$ regularly, for example should satisfy the Lipschitz condition with respect to the metric $d$ in $X$:
\begin{equation}\label{Lipschitz}
|\Phi(I,\Gamma)-\Phi(J,\Gamma)|\le C d(I,J).
\end{equation}
Similar discussion for the frequency of occurrence of fragments $I$ in databases (without taking into account the conformations of fragments) one can find in \cite{Nekrasov1}.

From the standard point of view conformations of polymers (in particular, fragments of proteins) belong to continuous space, and sequences of residues in polymers belong to discrete set. Coarse graining of conformations of fragments and the Lipschitz continuity property (\ref{Lipschitz}) shows that for fragments of proteins it is natural to consider the opposite assumption --- the conformations of fragments, up to small deformations, belong to small discrete set $Y_{\varepsilon}$, and the sequences (from the point of view of probabilities to take conformations in $Y_{\varepsilon}$) are quasi--continuous.

\section{Contacts of fragments}

It is natural to discuss the contribution to free energy of a protein from contacts of fragments. Let us consider amino acid residues with the numbers $i$ and $j$ in the protein, which are in contact (i.e. distance between the residues is sufficiently small) and the residues are not neighbors: $|i-j|>1$. The corresponding fragments with the centers in $i$, $j$ are $\Gamma_i$, $\Gamma_j$.

One has to distinguish the different kinds of contacts. Contacts in secondary structures are related to hydrogen bonds between the residues in the backbone of peptide chain. These contacts are not very specific (dependence on the types of amino acids is low) and their energy in some approximation can be taken proportional to the number of hydrogen bonds (i.e. the number of contacts).

Contacts between the different secondary structures and contacts in loops are made by side chains of amino acids and strongly depend on geometry and other properties of the corresponding fragments. Let us assume that the energy of a contact is determined by the contacting amino acid residues and their neighbors, i.e. by the contact of fragments of length three. If we would try to take into account longer contacting fragments it would be more complicated to find the statistics for contacts.

Thus we consider a contact of fragments (of length five) $(I,\Gamma_1)$ and $(J,\Gamma_2)$, contacting in the central (third) residues in the fragments, here $\Gamma_1,\Gamma_2\in Y_{\varepsilon}$. For these fragments we consider subfragments $(\widetilde{I},\widetilde{\Gamma}_1)$, $(\widetilde{J},\widetilde{\Gamma}_2)$ of length three (with the same centers), where $\widetilde{I}$, $\widetilde{J}$ are given by restriction of $I$, $J$ correspondingly. The conformation $\widetilde{\Gamma}$ can be considered as corresponding to a union of the sets $\Gamma\in Y_{\varepsilon}$ such that the restrictions of conformations of fragments in these sets to fragments of length three (with common central residue) have the distance between these restrictions of conformations less or equal to $\varepsilon$. The corresponding covering of the set of conformations of fragments of length three we denote $\widetilde{Y}_{\varepsilon}$.

Let us consider for a specific contact (i.e. contact between the different secondary structures or loops) the statistical potential
\begin{equation}\label{Psi}
\Psi(\widetilde{I},\widetilde{J},\widetilde{\Gamma}_1,\widetilde{\Gamma}_2)=-\log\left(\hbox{\rm frequency~of~occurrence}(\widetilde{I},\widetilde{J},\widetilde{\Gamma}_1,\widetilde{\Gamma}_2)\right),
\end{equation}
where we consider the frequency of occurrence of contacts of fragments ($\widetilde{I}$, $\widetilde{\Gamma}_1$), ($\widetilde{J}$, $\widetilde{\Gamma}_2$) where the centers of the fragments are in contact if the distance between the centers is less than some $\varepsilon'$.

Let us add to the functional (\ref{energy}) of free energy of a protein the term which describes the contacts of fragments
\begin{equation}\label{F_Phi_Psi}
F_1(I,\Gamma)=\lambda_1 \sum_{i,j\in C_1} 1 + \lambda_2 \sum_{i,j\in C_2}\Psi(\widetilde{I}_i,\widetilde{I}_j,\widetilde{\Gamma}_i,\widetilde{\Gamma}_j).
\end{equation}
Here $I$ is the sequence of amino acids in a protein, $\Gamma$ is the conformation of a protein; $I_i$, $\Gamma_i$ correspond to fragments of length five; $\widetilde{I}_i$, $\widetilde{\Gamma}_i$ correspond to fragments of length three; $C_1$ is the set of contacts in secondary structures (non specific contacts), $C_2$ is the set of specific contacts; in the sums the $i$-th and $j$-th residues are in contact; $\lambda_1,\lambda_2>0$ are some constants.

One can also take into account (lower) specificity of contacts in secondary structures (in particular beta sheets), substituting the summands in the first sum (over non specific contacts) in (\ref{F_Phi_Psi}) by the analogue of (\ref{Psi}), computed using some database. For example one can take into account hydrophobic interaction of side chains of amino acids.

\section{Hierarchy of structures in proteins}

In we present section we consider a generalization of the model of fragments (\ref{energy}), (\ref{F_Phi_Psi}) which takes into account a hierarchy of fragments of different lengths. In addition to the database $S$ of fragments we will use a database $T$ of proteins (this database should contain primary structures of proteins and their native conformations). We will understand conformations of proteins as sequences $(\Gamma_i)=\Gamma_1\dots\Gamma_{M}$ of coarse grained conformations of fragments. Thus a protein $(I,\Gamma)$ from $T$ generates  a sequence $(I_i,\Gamma_i)$ of fragments.

Let us consider for a protein $(I,\Gamma)\in T$ values $\Phi(I_i,\Gamma_i)$ of statistical potentials of fragments as a function of the number $i$ of the fragment. This function can be considered as a stepwise real valued function $\Phi(x)$ on the interval $[1/2,M+1/2]$, where the function at the interval $[i-1/2,i+1/2]$ equals to $\Phi(I_i,\Gamma_i)$.

Let us apply to this function the smoothing procedure by convolution with gaussian function $\frac{1}{\sigma\sqrt{2\pi}}e^{-\frac{x^2}{2\sigma^2}}$ with the variance $2\sigma$ ($2\sigma$ is close to one)
$$
f(x)=\int \Phi(y)\frac{1}{\sigma\sqrt{2\pi}}e^{-\frac{(x-y)^2}{2\sigma^2}}dy.
$$

Let us construct for any protein $(I,\Gamma)\in T$ a tree ${\cal T}(I,\Gamma)$ of ''hierarchy of basins'' \cite{Landscape} as follows. Let us fix some set $\{q_j\}$ of real numbers (the function $f(x)$ should take values which lie between some $q_l$ and $q_k$). This set can be taken ordered with respect to increasing of the indices, i.e. $q_i<q_j$ for $i<j$.

Let us consider in the interval $[1/2,M+1/2]$ the set $\{x: f(x)\le q_j\}$. This set is a union of intervals. These intervals are partially ordered by inclusion and the partial order is described by a tree, i.e. any two intervals either can be nonintersecting (modulo set of measure zero), or one of the intervals will contain the other (in this case the intervals will correspond to different $q_j$). The obtained tree we denote ${\cal T}(I,\Gamma)$ (this tree depends on the hierarchy $\{q_j\}$ of barriers). Vertices of the tree correspond to the intervals (''basins''), partial order in the tree is defined by inclusion of intervals, edges connect intervals embedded without intermediaries.

In the limiting case when the set $\{q_j\}$ is dense we will get the hierarchical partition of the interval $[1/2,M+1/2]$ by local maxima of the function $f(x)$, in general case we get a coarse grained hierarchical partition of $[1/2,M+1/2]$.

Similar procedure of hierarchical partition of proteins was discussed in papers by Nekrasov et. al. \cite{Nekrasov}, \cite{Nekrasov1}, \cite{Nekrasov2} (where the frequencies of occurrence for fragments $I_i$ were considered and the conformations $\Gamma_i$ of the fragments were not taken into account). It was shown that maximal branches in obtained trees correspond to domains in proteins.

Each interval from the described system of ''basins'' corresponds to some branch of the tree ${\cal T}(I,\Gamma)$. This branch contains a vertex corresponding to the interval and all vertices which are less with respect to the partial order (corresponding to subintervals of the given interval). We will denote the interval and the corresponding branch of the tree by $A$. For a given branch $A$ let us consider a set of integer points $i$ belonging to the interval $A$, and the corresponding set of fragments with conformations $\Gamma_i$. We will denote $(\Gamma_i)_{A}$ the obtained sequence of conformations of fragments.

Then, for all obtained in the described way sequences of fragments $(\Gamma_i)_A$ we compute the frequencies of occurrence in the database $T$ and the corresponding statistical potentials
$$
\Phi(A)=-\log \left(\hbox{frequency of occurrence of $(\Gamma_i)_A$ in $T$}\right).
$$

The idea of consideration of statistical potentials $\Phi(A)$ of branches of trees ${\cal T}(I,\Gamma)$ is based on the following observation: any branch of the trees under consideration (i.e. a sequence $(\Gamma_i)_A$) corresponds to some conformation which may occur in many proteins. In particular in \cite{Nekrasov}, \cite{Nekrasov1}, \cite{Nekrasov2} it was shown that maximal branches correspond to domains in proteins. Thus the conformational entropy of the hierarchical set of elements in proteins under discussion should be comparably low (in comparison to arbitrary possible sequences of conformations of fragments). Therefore we can use the database of conformations related to branches $A$ for investigation of conformational structures of proteins.

We will get a database of branches $A$ and corresponding conformations of fragments of proteins $(\Gamma_i)_A$. This database does not contain information about sequences of amino acids in the fragments since the entropy of sequences is large. 

Let us consider the contribution to the free energy of a protein $(I,\Gamma)$ of the form
\begin{equation}\label{F_2}
F_2(I,\Gamma)=\sum_{A:(\Gamma_i)_A\subset \Gamma}\lambda(A)\Phi(A).
\end{equation}
Here the summation runs over branches $A$ where the corresponding conformations are subsequences of neighbor fragments in $\Gamma=\Gamma_1\dots\Gamma_{M}$, i.e. $(\Gamma_i)_A\subset \Gamma_1\dots\Gamma_{M}$. There is no explicit dependence on the primary structure $I$ of the protein in the above expression (primary structure was used for the construction of the branches $A$ of the tree ${\cal T}(I,\Gamma)$).

Expression (\ref{F_2}) for the energy of conformation contains the information about selection of conformations of protein segments at the level of branches of the tree ${\cal T}(I,\Gamma)$. Positive constants $\lambda(A)$ should grow with increasing of length of the sequence $(\Gamma_i)_A$ (to compensate small values of the statistical potential $\Phi(A)$). It is natural to consider $\lambda(A)$ which depend only on the length of such sequence.
Thus the expression for $F_2$ means that conformations corresponding to hierarchically embedded space structures frequent in proteins are energetically profitable.

The total free energy of a protein in the model under consideration is a sum of contributions (\ref{energy}), (\ref{F_Phi_Psi}) and (\ref{F_2})
\begin{equation}\label{F}
F(I,\Gamma)=F_0(I,\Gamma)+F_1(I,\Gamma)+F_2(I,\Gamma).
\end{equation}

\medskip

\noindent{\bf Remark.}\quad The described above hierarchical markup of proteins is analogous to hierarchical syntax markup of texts. Text is a sequence of letters, which can be considered as containing several levels of hierarchy --- letters, words (combination of letters), collocations (combinations of words), phrases. Each level of hierarchy has lesser entropy in comparison to the set of arbitrary combinations of elements of previous level of hierarchy (for example, number of words is much less than the number of possible combinations of letters).

\medskip

\noindent{\bf Remark.}\quad Protein folding is a cooperative transition (similar to the first order phase transition). But in one dimensional systems there should be no phase transitions. One can consider the term (\ref{F_2}) in the model of the free energy of a protein as a description of this cooperative transition. This contribution is non--local, which may be compared to the existence of long range interactions in the Sherrington--Kirkpatrick model of spin glasses which predicts the glass transition (described by the hierarchical replica symmetry breaking method).

\section{Methods of analysis of protein conformations}

In the present section we discuss application of the introduced above model (\ref{F}) of statistical potentials of fragments to the investigation of protein conformations.

Let us discuss the problem of reproducing of the native conformation of a protein starting from its primary structure. We discuss two approaches. The first is based on minimization of the free energy functional (\ref{F}) over conformations $\Gamma$ (the problem of folding). The second is a variant of threading method where the scoring function for protein comparison is constructed with the help of functionals (\ref{energy}) and (\ref{F}).

\medskip

\noindent{\bf Problem of folding (combinatorial optimization).}\quad
The aim is, starting from the primary structure $I$ (sequence of amino acid residues in a protein), to reproduce the native conformation $\Gamma$ of a protein as the global minimum of the free energy functional (\ref{F}). We have to construct the sequence of conformations of fragments $\Gamma_i$ which minimizes (\ref{F}).

In this statement the problem of folding is a problem of combinatorial optimization. The condition of compatibility of neighbor fragments (namely, sparsity of the matrix $(C_{\Gamma\Delta})$ of intersections of fragments) considerably reduces the volume of the set for brute--force search for global minimum.

Complexity of this problem of combinatorial optimization can be reduced in the following way. Let us find fragments $I_i$ where the conformation is defined unambiguously by the sequence (i.e $\Phi(I_i,\Gamma_i)$ for such $I_i$ is concentrated on the unique $\Gamma_i$). We will call $I_i$ of this kind certain. Then for construction of the native conformation of a protein it is sufficient to minimize the functional of free energy on subsequences lying between two certain fragments. If the distance between two certain fragments is not very high then the problem of combinatorial optimization will not be very complex.

It was shown \cite{Nekrasov1} that short fragments with high occurrence frequency are situated sufficiently frequent along the protein sequence (long segments with low occurrence frequency of short fragments are very rare). From the point of view of the functional (\ref{F}) this results in reduction of complexity of the folding problem.

Problem of minimization of energy (\ref{F}) is the analogue of the maximal likelihood method in data analysis. Here the likelihood functional (the product of occurrence frequencies of conformations of fragments multiplied by the contributions of contacts of fragments and of the hierarchy of longer fragments) will be equal to exponent of the taken with opposite sign energy.

\medskip

\noindent{\bf Hierarchical approach to folding.}\quad
Let us discuss the problem of folding from the point of view of hierarchy of branches $A$ in (\ref{F_2}), (\ref{F}).
Search of minima of energy over conformations can be organized in different ways. Let us consider the following hierarchical algorithm: at the first step we search fragments $(I_i,\Gamma_i)$ with minimal energy (i.e. unambiguously foldable), then we try to include these fragments in different branches $(\Gamma_i)_A$, growing these branches by inclusion. The search will (in some approximation) be greedy --- at every step we have to minimize the energy for one branch (hierarchically increasing this branch), then we combine the obtained branches at the higher level of hierarchy of branches.  Thus we use the database of conformations related to branches $A$ for reducing of volume of the search.

In the literature \cite{Finkelstein} it was discussed that folding in real proteins is similar to the described above greedy procedure (with formation of nuclei of the native structure which grow in volume). Greediness of the search can considerably reduce volume of this search. This volume (for the described greedy search) can be estimated by the number of levels of hierarchy multiplied by the complexity of the search for one unit of hierarchy. This in principle can reduce the search from brute force enumeration of all conformations to the search over the set with the volume proportional to logarithm of the number of conformations. Similar point of view (hierarchy plus greedy search) is the basis of ''deep learning'' approach in machine learning, i.e. effective learning of multilayer hierarchical neural networks \cite{Hinton}, \cite{Bengio}.

Using this discussion, we conjecture that the Levinthal paradox (the problem of search for the global minimum of energy over exponentially large space of conformations) might be solved with the help of greedy search in the described hierarchical model of energy (\ref{F}) based on the data of bioinformatics.

\medskip

\noindent{\bf Threading.}\quad  In some approximation the majority of protein folds are known. Therefore native conformation of a protein can be found by comparison of a protein and proteins from some database with known native conformations.  The procedure of threading for recognition of conformation has the following form: a sequence of a protein is aligned with sequences of proteins with known conformations. Then the protein with better alignment can be used for modeling of the conformation of the protein under investigation.

Let us discuss a generalization of threading method based on the idea to take into account the statistics of short fragments of proteins.
We will compare a protein with the sequence $I$ (and unknown native conformation) with a protein $J$ from the database with conformation $\Gamma$ using alignment of $I$ and $J$.

Let us put the protein $I$ in conformation $\Gamma$ (as $J$) and consider the function $\Phi(I_i,\Gamma_i)$ as a function of the fragment $I_i$ --- the $i$-th fragment in $I$. Let us consider the similarity functional
\begin{equation}\label{PhiPhi}
r_{\Gamma}(I,J)=\sum_{i}\left|\Phi(I_i,\Gamma_i)-\Phi(J_i,\Gamma_i)\right|,
\end{equation}
where $J_i$ is the $i$-th fragment in $J$. This functional depends on the alignment of the two proteins and the corresponding alignment of fragments. One can consider a generalization of this functional by adding to (\ref{PhiPhi}) some terms which compare the contacts of fragments of the form (\ref{Psi}).

Two proteins $I$ and $J$ are similar if local physical properties of these proteins in conformation $\Gamma$ (the native conformation of $J$) are similar with respect to the functional of free energy (\ref{energy}). In this case the value of functional (\ref{PhiPhi}) will be low and one can assume that these proteins have the same native conformation.

One of generalizations of threading method is to consider threading not of a single protein but a family of threadings of homologs and take some averaging \cite{Rigden}. This approach has some similarity with the described above where we take into account the statistics of fragments averaged over some neighborhood of fragments in the primary structure $I$.

\section{Structural alignment and model of fragments}

Application of statistical potentials for fragments allows us to modify not only threading procedure but also alignment construction. In the present section we introduce a version of structural alignment construction related to the model of fragments. Structural alignment is a modification of alignment of proteins which takes in account the structures of proteins (information about the conformations).

Let us recall the definition of alignment, see for example \cite{Pevzner}. Edit distance between two sequences of symbols from finite alphabet is the minimal number of edit operations which map one sequence to the other. Set of edit operations usually contain insertions, deletions and substitutions of symbols. One can consider global alignment of sequences and local alignment (of segments of sequences).

\bigskip

\noindent{\bf Definition of alignment.}\quad Let ${\cal A}$ be a $k$-letter alphabet, $V$ and $W$ be finite sequences of symbols from ${\cal A}$. Let ${\cal A}'={\cal A}\bigcup \{-\}$ be extended alphabet where the symbol $\{-\}$ is called the gap symbol.

Alignment of two sequences $V=v_1\dots v_n$ and $W=w_1\dots w_m$ is a matrix with two lines of equal length $l\ge n,m$, the first line of the matrix contains a sequence $\widetilde{V}=\widetilde{v}_1\dots \widetilde{v}_l$ obtained from $V$ by insertion of $l-n$ gaps in some order, the second line contains $\widetilde{W}=\widetilde{w}_1\dots \widetilde{w}_l$ obtained from $W$ by insertion of $l-m$ gaps. Columns with two gaps are forbidden.

Columns of the alignment matrix which contain gaps at the first line are called insertions, columns containing gaps at the second line are deletions. Columns containing identical symbols in both lines are called matches, and columns containing different symbols are called mismatches.

One puts in correspondence to each column of the alignment matrix the score -- a real number depending on the symbols in the column. Score of the alignment is a sum over columns of the matrix:
\begin{equation}\label{weight}
\delta(\widetilde{V},\widetilde{W})=\sum_{i=1}^{l}\delta(\widetilde{v}_i,\widetilde{w}_i).
\end{equation}

The alignment problem is to find alignment with maximal score.

Alignment ($\widetilde{V}$, $\widetilde{W}$) of sequences $V$, $W$ corresponds to a sequence of edit operations of the line $V$ which map $V$ to $W$. Edit operations correspond to columns of the alignment matrix ($\widetilde{V}$, $\widetilde{W}$) and can be performed in arbitrary order. The operations are as follows: insertion of gap in the line $V$ (at the position corresponding to the column), deletion of a symbol in $V$, mismatch corresponds to substitution of a symbol from $V$ by corresponding symbol from $W$.

In the simplest case one can put scores equal to $\delta(x,x)=1$ for matches, $\delta(x,y)=-\mu$ for mismatches and $\delta(x,-)=\delta(-,x)=-\sigma$ for insertions or deletions:
$$
{\rm score~}({\rm ~alignment~})=\#({\rm ~matches~})-\mu \#({\rm ~mismatches~})-\sigma \#({\rm ~indels~}).
$$

For alignment of proteins one can use PAM or BLOSUM matrices for scores of amino acid substitutions, i.e. the score for substitutions of amino acids  $i$ and $j$ will be a function of the matrix element $A_{ij}$.

\bigskip

\noindent{\bf Structural alignment.}\quad Let us consider alignment ($\widetilde{V}$, $\widetilde{W}$) of sequences $V=v_1\dots v_n$ and $W=w_1\dots w_m$ (sequences of amino acids in some proteins). Let the conformation $\Gamma$ of the protein $W$ be known, and let us consider the corresponding sequence of conformations of fragments (of length five)  $\Gamma=\Gamma_3\Gamma_4\dots\Gamma_{m-2}$ (here we enumerate fragments by numbers of amino acids in central positions of the fragments).

In the aligned sequences $\widetilde{V}$, $\widetilde{W}$ one can consider fragments $\widetilde{V}_i$, $\widetilde{W}_i$ of length five with centers in the $i$-th positions. We consider only the case when both fragments $\widetilde{V}_i$, $\widetilde{W}_i$ do not contain gaps. Then, let $\Gamma_i$ be a conformation of the fragment $\widetilde{W}_i$ (obtained by restriction of the conformation of the protein $(W,\Gamma)$).

Let us consider statistical potentials $\Phi\left(\widetilde{V}_i,\Gamma_i\right)$, $\Phi\left(\widetilde{W}_i,\Gamma_i\right)$ and define the structural score for these fragments by the formula
\begin{equation}\label{structural}
\delta\left(\widetilde{V}_i,\widetilde{W}_i,\Gamma_i\right)= |\Phi\left(\widetilde{V}_i,\Gamma_i\right)-\Phi\left(\widetilde{W}_i,\Gamma_i\right)|.
\end{equation}

The score $\delta\left(\widetilde{V}_i,\widetilde{W}_i,\Gamma_i\right)$ is low when the statistical potentials of fragments $\widetilde{V}_i$, $\widetilde{W}_i$ in the conformation $\Gamma_i$ are similar. These structural scores define contributions to the score functional (\ref{weight}) of the alignment. Let us note that here instead of maximization we consider minimization of the score of the alignment (i.e. the lower score is the better).

Another case of the construction under consideration is the local alignment of segments in sequences $V$ and $W$ without gaps. The alignment score takes the form
\begin{equation}\label{max}
\sum_i |\Phi\left(V_i,\Gamma_i\right)-\Phi\left(W_i,\Gamma_i\right)|,
\end{equation}
i.e. we align segments of equal length without gaps in sequences $V$, $W$ and consider the corresponding alignment of fragments $V_i$, $(W_i,\Gamma_i)$ (of length five) lying in the segments.  The alignment score will be good (low) when statistical potentials of the aligned fragments in the same conformations will be similar. 

Both functionals (\ref{structural}), (\ref{max}) are nonlocal with respect to sequences $V$, $W$. Thus there are no algorithms of dynamical programming for finding extremes of these functionals. One can construct local alignment of short segments and then elongate these segments (as in the BLAST algorithm) taking into account the structural alignment score.

\section{Application to protein dynamics}

In the discussed above approach we put in correspondence to a protein, in addition to the primary structure and the native conformation, another two sequences --- a sequence $(I_i)$ of fragments and a sequence $(\Gamma_i)$ of coarse grained conformations of fragments; we consider two sets of contacts between fragments (non--specific and specific); and two functions $\Phi$ and $\Psi$ (frequencies of occurrence of fragments and of specific contacts of fragments). We also consider the hierarchy of ''basins'' for the function $\Phi$ and the set of statistical potentials for this hierarchy. We use this data to construct functional (\ref{F}) of free energy and discuss application of this functional to investigation of proteins.

The function $\Phi$ has the meaning of binding energy of a fragment.
It was shown \cite{Nekrasov1}, \cite{Nekrasov2} that similar function identifies protein domains as segments between deep local minima of $-\Phi$ (taken as a function of $i$-th fragment), in this paper only statistics of sequences of amino acids in fragments was used and conformations were not taken into account. Parts of a protein with high values of $-\Phi$ have high binding energy and therefore high rigidity, segments with low rigidity separate the domains. This approach can be used for description of mechanical properties of a protein globule.

Expression (\ref{F}) for free energy describes the energy in a fixed conformation (native state). For description of relaxation dynamics at small distance one can assume that any  fragment $I_i$ of the protein has fixed coarse grained conformation $\Gamma_i\in Y_{\varepsilon}$ which does not change in the process of dynamics. Some fragments may deform but since conformations $\Gamma_i\in Y_{\varepsilon}$ are defined up to small perturbation $\varepsilon$ one can assume that classes of conformations in $Y_{\varepsilon}$ remain unchanged. Thus coarse grained description does not describe the dynamics.

Protein dynamics can be described by a model of spring with variable elasticity. The spring will describe the backbone of the peptide chain. Elasticity of the spring will be given by the function $-\Phi$: elasticity at the $i$-th residue will be equal to $-\Phi(I_i,\Gamma_i)$ (we may add a constant to the function $-\Phi(I_i,\Gamma_i)$ to make elasticity  positive). Also one can take into account adhesion of fragments described by the function $\Psi$.

The function $-\Phi$ (considered as a function of the number of fragment $i$) has the described above hierarchical form   --- graph of the function has the form of a hierarchy of elevations separated by local minima of hierarchical depth. This observation may be compared with the concept of molecular machine as a realization of hierarchical (fractal, or crumpled) globule \cite{AvetisovNechaev,AvetisovNechaev1}, see also \cite{Smrek} for discussion of relation of fractal globules, space filling curves and DNA conformation in chromosomes.
The discussed in the present paper approach differs from the hierarchical version of the model of elastic networks \cite{Togashi} considered in \cite{AvetisovNechaev,AvetisovNechaev1}. First, we consider the model of spring with variable elasticity instead of the model of elastic network, second, the origin of hierarchy in the model is the hierarchical structure of the function $\Phi$ instead of a hierarchical structure of the protein globule.

Hierarchy of ''basins'' for the function $\Phi$ (branches $(\Gamma_i)_A$ of the tree ${\cal T}(I,\Gamma)$ of basins) can be considered as a description of the construction of the molecular machine --- the deeper the branch the higher the energy of deformation of this branch, one branch can move relative to another as a whole.

\section{Conclusion}

In the present paper we construct a model of free energy of proteins based on bioinformatics. Using statistics of short fragments in proteins, we build a family of statistical potentials which describe joint distribution of sequences of amino acids in the fragments and conformations of the fragments.

The important feature of the model is that all the data (sequences and conformations) in the model are discrete. In particular, a conformation of a protein is represented as a sequence of conformations of fragments, which belong to the set of small size. This kind of representation allows us to use data of bioinformatics (which usually have a form of sequences of some symbols) on a par with physical data. Possibility of this representation is granted by the smallness of $\varepsilon$--entropy of the space of conformations of fragments (known from the experiments).

In the model under consideration the free energy of a protein is a sum of statistical potentials of short fragments of a protein, contacts of fragments, and hierarchical family of longer fragments. In general, this model is example of physical model of a complex system based on big data (in our case on the data of bioinformatics).

We also discuss application of the considered model of statistical potentials for protein fragments to structural alignment of proteins. Modification of the scoring functional for alignment which takes into account statistical potentials for fragments is considered and application to threading procedure is discussed.

\bigskip

\noindent{\bf Acknowledgments.}\quad This work is supported by the RSF under a grant 14--11--00687 and performed in the Steklov Mathematical Institute.

\end{document}